
\magnification = 1200
\font\eightrm=cmr8
\font\eighti=cmmi8
\font\eightsy=cmsy8
\font\eightbf=cmbx8
\font\eighttt=cmtt8
\font\eightit=cmti8
\font\eightsl=cmsl8
\font\sixrm=cmr6
\font\sixi=cmmi6
\font\sixsy=cmsy6
\font\sixbf=cmbx6
\catcode`@11
\newskip\ttglue

\def\eightpoint{\def\rm{\fam0\eightrm}
\textfont0=\eightrm \scriptfont0=\sixrm \scriptscriptfont0=\fiverm
\textfont1=\eighti \scriptfont1=\sixi \scriptscriptfont1=\fivei
\textfont2=\eightsy \scriptfont2=\sixsy \scriptscriptfont2=\fivesy
\textfont3=\tenex \scriptfont3=\tenex \scriptscriptfont3=\tenex
\textfont\itfam=\eightit \def\it{\fam\itfam\eightit}
\textfont\slfam=\eightsl \def\sl{\fam\slfam\eightsl}
\textfont\ttfam=\eighttt \def\tt{\fam\ttfam\eighttt}
\textfont\bffam=\eightbf
\scriptfont\bffam=\sixbf
\scriptscriptfont\bffam=\fivebf \def\bf{\fam\bffam\eightbf}
\tt \ttglue=.5em plus.25em minus.15em
\normalbaselineskip=6pt
\setbox\strutbox=\hbox{\vrule height7pt width0pt depth2pt}
\let\sc=\sixrm \let\big=\eightbig \normalbaselines\rm}
\newinsert\footins
\def\newfoot#1{\let\@sf\empty
  \ifhmode\edef\@sf{\spacefactor\the\spacefactor}\fi
  #1\@sf\vfootnote{#1}}
\def\vfootnote#1{\insert\footins\bgroup\eightpoint
  \interlinepenalty\interfootnotelinepenalty
  \splittopskip\ht\strutbox 
  \splitmaxdepth\dp\strutbox \floatingpenalty\@MM
  \leftskip\z@skip \rightskip\z@skip
  \textindent{#1}\footstrut\futurelet\next\fo@t}
\def\fo@t{\ifcat\bgroup\noexpand\next \let\next\f@@t
  \else\let\next\f@t\fi \next}
\def\f@@t{\bgroup\aftergroup\@foot\let\next}
\def\f@t#1{#1\@foot}
\def\@foot{\strut\egroup}
\def\footstrut{\vbox to\splittopskip{}}
\skip\footins=\bigskipamount 
\count\footins=1000 
\dimen\footins=8in 

\def\ref#1{$^{#1}$}
\def\flex{\raise 6pt\hbox{$\leftrightarrow $}\! \! \! \! \! \! }
\def\oversome#1{ \raise 8pt\hbox{$\scriptscriptstyle #1$}\! \! \! \! \! \! }

\newbox\bigstrutbox
\setbox\bigstrutbox=\hbox{\vrule height10pt depth5pt width0pt}
\def\bigstrut{\relax\ifmmode\copy\bigstrutbox\else\unhcopy\bigstrutbox\fi}
\def\refer[#1/#2]{ \item{#1} {{#2}} }
\def\rev<#1/#2/#3/#4>{{\it #1\/} {\bf#2}, {#3}({#4})}
\def\boxit#1{\vbox{\hrule\hbox{\vrule\kern3pt
\vbox{\kern3pt#1\kern3pt}\kern3pt\vrule}\hrule}}

\def\2figure#1#2#3#4{\vbox{ \hrule width#1truecm \hbox{\vrule height#2truecm
\hskip #1truecm
\vrule height#2truecm }\hrule width#1truecm \hbox{\vrule\vbox{\hsize #1truecm
\baselineskip=10pt
\noindent\strut#3}\vrule}\hrule width#1truecm
\hbox{\vrule\vbox{\hsize #1truecm
\baselineskip=10pt
\noindent\strut#4}\vrule}\hrule width#1truecm  }}
\def\3figure#1#2#3#4#5{\vbox{ \hrule width#1truecm \hbox{\vrule height#2truecm
\hskip #1truecm
\vrule height#2truecm }\hrule width#1truecm \hbox{\vrule\vbox{\hsize #1truecm
\baselineskip=10pt
\noindent\strut#3}\vrule}\hrule width#1truecm
 \hbox{\vrule\vbox{\hsize #1truecm
\baselineskip=10pt
\noindent\strut#4}\vrule}
\hrule width#1truecm \hbox{\vrule\vbox{\hsize #1truecm
\baselineskip=10pt
\noindent\strut#5}\vrule}\hrule width#1truecm  }}

\def\sqr#1#2{{\vcenter{\hrule height.#2pt
   \hbox{\vrule width.#2pt height#1pt \kern#1pt
    \vrule width.#2pt}
    \hrule height.#2pt}}}


\def\smin{\,\raise 0.06em \hbox{${\scriptstyle \in}$}\,}
\def\smsubset{\,\raise 0.06em \hbox{${\scriptstyle \subset}$}\,}

\def\Natural{\hbox{\hskip 1.5pt\hbox to 0pt{\hskip -2pt I\hss}N}}

\def\Rational{\hbox{\hbox to 0pt{\hskip 2.7pt \vrule height 6.5pt
                                  depth -0.2pt width 0.8pt \hss}Q}}
\def\Real{\hbox{\hskip 1.5pt\hbox to 0pt{\hskip -2pt I\hss}R}}
\def\Complex{\hbox{\hbox to 0pt{\hskip 2.7pt \vrule height 6.5pt
                                  depth -0.2pt width 0.8pt \hss}C}}

\nopagenumbers
\vfill\eject
\centerline {\bf String inspired effective Lagrangian and}
\centerline {\bf Inflationary Universe}
\vskip 1.5cm
\centerline {\bf E. Abdalla\newfoot{\ref\dagger}{Work partially supported by
CNPq.} and A.C.V.V. de Siqueira\newfoot{\ref *}{Work supported by CAPES;
permanent address, Universidade Cat\'olica de Recife, Pernambuco, Brazil.}}
\centerline { Instituto de F\'{\i}sica da Universidade de S\~ao Paulo}
\vskip 2cm
\centerline {\bf Abstract}

\noindent We consider a string inspired effective Lagrangian for the graviton
and dilaton, containing Einstein gravity at the zero slope limit. The numerical
solution of the problem shows asymptotically an inflationary universe. The time
is measured by the dilaton, as one expects. The result is independent of the
introduction of ad-hoc self interactions for the dilaton field.
\vskip.5cm

\hfill Universidade de S\~ao Paulo\quad

\hfill IFUSP-preprint-1028\phantom{Paulo}\quad

\hfill January 1993\phantom{Paulo}\quad

\vfill\eject

\countdef\pageno=0 \pageno=1
\newtoks\footline \footline={\hss\tenrm\folio\hss}
\def\folio{\ifnum\pageno<0 \romannumeral-\pageno \else\number\pageno \fi}
\def\advancepageno{\ifnum\pageno<0 \global\advance\pageno by -1
\else\global\advance\pageno by 1 \fi}

The inflationary scenario solves important problems posed by the big bang
cosmolo\-gy\ref{1,2,3}. Indeed, the observed isotropy of the cosmic background
radiation, as well as the unstable value $\Omega\sim 1$ of the observed ration
of the density of the universe and the critical density can only be explained
by the exponential growth of the radius of the universe\ref{2,3} at an early
stage. This is known as the inflationary scenario and the way to achieve this
description is by means of the introduction of the Coleman-Weinberg
potential\ref{2,4,5} which develops a false vacuum as the primordial
temperature lowers. The decay of that metastable state develops the required
growth of the radius.

We aim at verifying whether a string inspired cosmological model could develop
that kind of behavior. In fact, a modified Einstein equation might lead to such
a behavior\ref{6}. In fact, study of strings moving on a background lead  to
new  equations for the gravitational fields, with quantum corrections for the
Einstein equations, as shown in [7] (see also [8],[9]). Due to the way the
$\alpha$ - parameter appears in the solution, it is convenient to start the
discussion by the effective action obtained as a consequence of the
requirement of conformal invariance of string in a background field\ref{7,8}
$$S=\int d^4x\sqrt{-g}e^{-2\phi}\left[ R^\mu_{\;\mu} +4\left(
\nabla\phi\right)^2+{1\over 4}\alpha
R_{\mu\nu\rho\sigma}R^{\mu\nu\rho\sigma}\right]\eqno(1)$$

The field equations derived from eq.(1) are obtained after some tedious but
simple algebra, and read
$$\eqalignno{\beta_{\mu\nu}^g&=R_{\mu\nu}+2\nabla_\mu\nabla_\nu\phi
+{1\over 2}\alpha
R_{\mu\beta\gamma\delta}R_\nu^{\phantom{\nu}\beta\gamma\delta}\cr
&+{1\over 2}\alpha
\left[ 2\nabla_\gamma\nabla_\pi\phi -4\nabla_\gamma\phi\nabla_\pi\phi
+4\nabla_\gamma\phi\nabla_\pi -\nabla_\gamma\nabla_\pi\right]
\left(R_{\mu\phantom{\gamma\pi}\nu}
^{\phantom{\mu}\gamma\pi\phantom{\nu}}+
R_{\nu\phantom{\gamma\pi}\mu}
^{\phantom{\nu}\gamma\pi\phantom{\mu}}\right)=0 &(2a)\cr
\beta^\phi&=\nabla^2\phi-\left(\nabla\phi\right)^2+{1\over 4}R^\mu_\mu +{1\over
16}\alpha R_{\mu\nu\rho\sigma}R^{\mu\nu\rho\sigma} =0 &(2b)\cr}$$

We wish to obtain the consequences of equations (2) for the big bang cosmology.
Therefore, we consider a Robertson-Walker type metric, defined by the line
element $ds^2$ as
$$ds^2=d{x^0}^2-R(t)^2\left[ {dr^2\over 1-kr^2}+r^2d\omega^2\right]\quad
.\eqno(3) $$

Using the above metric, it is not difficult to go back to (2) and obtain the
resulting differential equations
$$\eqalign{3{{\ddot R}\over R}-2{\partial^2\phi\over\partial
t^2}-3\alpha\!\!\left(\!{\ddot R\over R}\!\right)^2\!\!-\! 6\alpha
{\ddot R\dot R\over R^2}{\partial\phi\over\partial t}
+3\alpha{\dot R\over R}\!\left[
{d\over dt}\!\left(\!{\ddot R\over R}\!\right)\!+2 {\dot R\over R}\!
\left(\!{\ddot R\over R}\! -\!{k+\dot R^2\over R^2}\!\right)\!\right]=0\cr
\ddot\phi +3{\dot R\over R}\dot \phi -\dot \phi^2- {3\over 2}\left[ {\ddot
R\over R}+{k+\dot R^2\over R^2}\right] +{3\over 2}\alpha
\left[ \left({\ddot R\over
R}\right)^2 +\left({k+\dot R^2\over R^2}\right)^2\right] =0\cr}\eqno(4)$$

We have to solve these (highly non linear) equations. An analytical solution of
the above system was beyond our abilities. However, we could easily obtain a
numerical solution. We first chose unities such that $\alpha=1$. We also
allowed for a dilaton effective self interaction\ref{9}
$$V(\phi)={1\over 2}m^2\phi^2 + {1\over 4}g\phi^4\quad .\eqno(5)$$

However, the results were not much sensitive to the above constants. We have
obtained results as follows. In general there is a transient region, for small
values of the time, where the solution depends very much on the initial
conditions. However, one quickly runs into the asymptotic region, where the
radius grows exponentially, and the dilaton is linear on time. Some case
examples are shown in figure 1. We have selected several initial conditions in
order to illustrate the fact that the result is quite robust.

This is generally the case for $k=1$. For $k=0,-1$ the transient region
persisted for a longer time, and we could not arrive to it in most cases as
shown in examples in fig. 2, although for $k=0$ we expect the same behavior as
for $k=1$ on the ground of the arguments shown below. Once
the asymptotic region has been reached, it is not difficult to see that the
Hubble constant defined in
$$ R=R_0e^{\chi t}\eqno(6)$$
and the relation between time and the dilaton
$$\dot\phi=\xi=\dot\phi (t=0)\eqno(7)$$
are uniquely determined in terms of the string constant. Indeed, equations (4)
imply (7) once (6) has been used, in the asymptotic region ($t>>1/\chi$). The
constants $\chi$ and $\xi$ are solutions to the algebraic equations
$$\eqalign{\alpha\chi^2+2\xi\alpha\chi -1&=0\cr
\xi^2 -3\chi\xi (1-\alpha\chi^2)+{3\over
2}\chi^2&=0.\cr}\eqno(8) $$

In principle it is possible to obtain $\chi<0$, but this solution is
incompatible with the approximations made (in fact, we expect that case to
describe a shrinking universe, although this claim is rather speculative; in an
oversimplified model shown below, in eq. (11), this is the case).
Thus we obtain
$$\chi=-\xi +\sqrt{\xi^2+{1\over \alpha}}.\eqno(9)$$

Therefore we arrive also at an algebraic equation for $\xi$:
$$12\alpha\xi^4+2\xi^2-{3\over
2\alpha}+3(1-4\alpha\xi^2)\xi \sqrt{\xi^2+{1\over
\alpha}} =0\eqno(10)$$

It is easy to verify that $\xi=\tilde\xi/\sqrt{\alpha}$
where $\tilde\xi$ obeys a $\alpha$-independent equation. For infinite
string tension ($\alpha\to\infty$), $\xi\to 0$. It is interesting to
verify that the oversimplified equation for the graviton
$$R_{\mu\nu} +{1\over 2}
\alpha R_{\mu\rho\pi\sigma}R_{\nu}^{\phantom{\nu}\rho\pi\sigma}=0
\eqno(11)$$
has also a solution of type (3),(6), with $\chi=\pm 1/\sqrt{\alpha}$. Indeed,
the large distance behavior is governed by the above equation. The doubling of
the sign is also noteworthy, since we have not only the case of an inflationary
universe, but also a compactified type dimension. Thus all richness seeked as
desired properties of string theories seem to show up already in these
oversimplified models.

Some numerical examples are worthwhile noticing. First, the results are
independent of the parameters $m^2$ and $g$, unless they are higher than 100,
showing that the result is very robust. In order to freeze the dilaton field,
we need $m^2,g\sim 200$. Moreover, for $k=1$, the result is also very regular.
We arrive at $\xi \sim - .5$, and $\chi=1.5$.
However, for $k=0,-1$ the results show instability, as we have shown in fig.2.

Further details as well as inclusion of further modes, such as antisymmetric
tensor fields as well as supersymmetry inspired fields are under
investigation\ref{10}.
\vskip 1cm

\centerline {\bf Figure captions.}

\item{} Figure 1: diagrams showing the time dependence of $\dot\phi$ $\ln R$,
and $\ln \dot R $ for $k=1$. The numbers above the diagram concern $\alpha$,
$k$, $m^2$, $g$, $\phi_0$, $\dot\phi_0$, $R_0$, $\dot R_0$, $\ddot R_0$,
respectively; Notice that we chose $\alpha$ to be always unity.

\item{} Figure 2: diagrams obtained for $k=0$.

\vskip 2cm
\centerline {\bf References}

\refer[[1]/A.H. Guth, Phys. Rev. {\bf D23} (1981)347-356.]

\refer[[2]/A.D. Linde, Phys. Lett. {\bf 108B} (1982)389-393.]

\refer[[3]/A. Zee, ``Unity of Forces in the Universe", World Scientific
Publishing, 1982.]
\refer[/L.F. Abbott, So-Young Pi, ``Inflationary Cosmology", World
Scientific Publishing, 1986.]

\refer[[4]/A.D. Linde, Rep. Prog. Phys. {\bf 42},389(1979)]

\refer[[5]/S. Coleman and E. Weinberg, Phys. Rev. {\bf D7} (1973)1888.]

\refer[[6]/A. Starobinski, Phys. Lett. {\bf 91B} (1980)99-102.]

\refer[[7]/C.G. Callan, I.R. Klebanov and M.J. Perry, Nucl. Phys. {\bf B278}
(1986)78.]

\refer[/C.G. Callan, E.J. Martinec, M.J. Perry and D. Friedan, Nucl. Phys.
{\bf B262} (1985)593.]

\refer[/A.A. Tseytlin, Class. Q. Grav.{\bf 9} (1992)979.]

\refer[[8]/M.B. Green, J.H. Schwarz and E. Witten, ``Superstring Theory",
Cambridge University Press, 1987.]

\refer[[9]/A.A. Tseytlin, Proceeding of the International  Workshop on
String, Quantum Gravity and Physics at the Plank Energy Scale, Erice, 1992,
DAMTP-92-36. ]

\refer[[10]/E. Abdalla, A.C.V.V. Siqueira, to appear.]

\end